\documentstyle[preprint,aps,epsfig]{revtex}

\def\be{\begin{equation}}
\def\de{\end{equation}}
\def\te{\tilde{\epsilon}}
\def\ep{\epsilon}

\begin{document}
\draft
\title{Electrorotation of colloidal suspensions}
\author{J. P. Huang$^1$, K. W. Yu$^1$ and G. Q. Gu$^{1,2}$}
\address{$^1$Department of Physics, The Chinese University of Hong Kong,
 Shatin, NT, Hong Kong}
\address{$^2$College of Information Science and Technology, 
 East China Normal University, \\ Shanghai 200 062, China}
\maketitle

\begin{abstract}
When a strong electric field is applied to a colloidal suspension, 
it may cause an aggregation of the suspended particles in response to the 
field. In the case of a rotating field, the electrorotation (ER) spectrum 
can be modified further due to the local field effects arising from the 
many-particle system. To capture the local field effect, we invoke the 
Maxwell-Garnett approximation for the dielectric response. 
The hydrodynamic interactions between the suspended particles can also 
modify the spin friction, which is a key to determine the angular 
velocity of ER. By invoking the spectral representation approach, 
we derive the analytic expressions for the characteristic frequency at 
which the maximum angular velocity of ER occurs. From the numerical 
caculation, we find that there exist two sub-dispersions in the ER 
spectrum. However, the two characteristic frequencies are so close that 
the two peaks actually overlap and become a single broad peak.
We report a detailed investigation of the dependence of the 
characteristic frequency and the dispersion strength of ER on various 
material parameters. 
\end{abstract}
\vskip 5mm

\section{Introduction}

AC electrokinetic pheonomena, such as dielectrophoresis and 
electrorotation 
(ER) have been investigated to yield the selective manipulation and 
characterization of biological cells.
Under the action of external fields, colloidal particles or biological 
cells in suspensions exhibit rich fluid-dynamic behavior as well as 
dielectric response. It is also interesting to investigate their 
frequency-dependent responses to ac fields, which leads to valuable 
information on the structural (Maxwell-Wagner) polarization effects 
\cite{Gimsa,Gimsa99}. 
The polarization is characterized by a variety of characteristic 
frequency-dependent changes known as dielectric dispersion.

In the last two decades, various experimental tools have been developed 
to 
analyze the polarization of biological cells - dielectric spectroscopy 
\cite{Asami80}, dielectrophoresis \cite{Fuhr} and ER \cite{Gimsa91} 
techniques. Among these techniques, conventional dielectrophoresis and ER 
are usually applied to analyze the frequency dependence of translations 
and rotations of cells in a rotating electric field, respectively 
\cite{Fuhr,Gimsa91}. Moreover, one is able to monitor the cell movements 
with the aid of automated video analysis \cite{Gasperis} as well as light 
scattering methods \cite{Gimsa99}.
In ER, a dipole moment is induced in response to the rotating field. 
Then any dispersion process is able to cause a phase shift between the 
induced dipole moment and the external field vector, resulting in a 
desired 
torque which causes the cells to rotate.

In the dilute limit, the ER of individual cell can be predicted by 
ignoring the mutual interaction between cells. However, the cells may be 
aggregated under the influence of the external field. 
In this case the Brownian motion can be neglected, and thus the system
becomes non-dilute even though it is initially dilute.
Then the role of mutual interaction will become dominant. 
As an initial model, we studied the ER of two approaching spherical 
particles in the presence of a rotating electric field in a recent paper 
\cite{Huang2}. We showed that when the two particles approach and finally 
touch, the mutual polarization interaction between the particles leads to 
a 
change in the dipole moment of individual particles and hence the ER 
spectrum, as compared to that of isolated particles. The mutual 
polarization effects were captured via the multiple image method 
\cite{Wan}. From the results, we found that the mutual polarization 
effects can lower the characteristic frequency at which the maximum ER 
angular velocity occurs.

When the volume fraction of the suspension becomes large, the mutual 
interactions between the suspended particles can also modify the spin 
friction, which is a key to determine the angular velocity of ER. 
For ER of two particles in a rotating electric field, we have 
successfully 
applied the spectral representation to calculate the dispersion 
frequency. 
However, the determination of the spin friction was lacking.

For two particles, the basic tool is the reflection method \cite{Happel}, 
being analogous to the multiple image method in electrostatics, but being 
valid for two particles only. For more than two particles, we need a 
first-principles method, e.g., the Green's function (Oseen tensor) 
formulation. For a dilute suspension, however, one may adopt the 
effective medium theories \cite{Choy} to capture the effective viscosity 
of a suspension. Thus, in this case, the ER spectrum can be modified further.

Regarding the spectral representation approach \cite{Bergman}, 
it is a rigorous mathematical formalism of the effective dielectric 
constant of a two-phase composite material.
It offers the advantage of the separation of material parameters 
(namely the dielectric constant and conductivity) from the particle
structure information, thus simplifying the study. 
In the present work, we will derive the analytic expressions for the 
characteristic frequencies by using the spectral representation approach.

\section{Formalism}

We consider homogeneous, spherical particles exposed to a rotating 
electric 
field of frequency $f$, in which particles of complex dielectric constant 
$\tilde{\epsilon}_1=\epsilon_1+\sigma_1/(i2\pi f)$ are 
dispersed in a suspension of 
$\tilde{\epsilon}_2=\epsilon_2+\sigma_2/(i2\pi f)$, with $i=\sqrt{-1}$. 
Here, $\epsilon_i$ and $\sigma_i$ represent the dielectric constant and 
conductivity, respectively. 

For the effective dielectric constant $\te_e$ of the whole system, we have:
\be
{\te_e - \te_2 \over \te_e + 2 \te_2} = p {\te_1 - \te_2 \over \te_1 + 2 
\te_2},
\label{mga}
\de
where $p$ is the volume fraction of partices. Note that 
the $p=0$ limit throughout the paper is just the isolated spherical 
result \cite{Huang2}. In this case, the dipole factor of the particle
is given by
\be
b = {\te_1 - \te_e \over \te_1 + 2 \te_e}.
\label{b}
\de

In order to obtain the analytic expression for the characteristic 
frequency, we resort to the spectral representation approach.
By introducing a material parameter 
\begin{eqnarray}
\tilde{s}&=&\left(1-\frac{\te_1}{\te_2}\right)^{-1},\nonumber
\end{eqnarray}
Eq.(\ref{b}) admits the form:

\be
b=\frac{F_1}{\tilde{s}-s_1}+\frac{F_2}{\tilde{s}-s_2},
\label{repb}
\de
where
\begin{eqnarray}
F_1&=&-\frac{(1-p)(\sqrt{8+p}-\sqrt{p})}{6\sqrt{8+p}},\ \ \
F_2=-\frac{(1-p)(\sqrt{8+p}+\sqrt{p})}{6\sqrt{8+p}},\nonumber\\
s_1&=&\frac{1}{6}\left(2+p-\sqrt{p(8+p)}\right),\ \ \ \
s_2=\frac{1}{6}\left(2+p+\sqrt{p(8+p)}\right).\nonumber
\end{eqnarray}

After some simple manipulations, Eq.(\ref{repb}) becomes \cite{Huang2,Huang}
\be
b=\frac{F_1}{s-s_1}+\frac{\Delta \ep_1}{1+if/f_{c1}}
 +\frac{F_2}{s-s_2}+\frac{\Delta \ep_2}{1+if/f_{c2}},
\label{repb2}
\de
where the dispersion magnitudes $\Delta\ep$ and characteristic 
frequencies $f_{c}$ being given by
\begin{eqnarray}
\Delta\ep_1&=&F_1\frac{s-t}{(t-s_1)(s-s_1)},\ \ \ 
\Delta\ep_2=F_2\frac{s-t}{(t-s_2)(s-s_2)}, \nonumber\\
f_{c1}&=&{1\over 2\pi}\frac{\sigma_2 s(t-s_1)}{\epsilon_2 t(s-s_1)},\ \ \
f_{c2}={1\over 2\pi}\frac{\sigma_2 s(t-s_2)}{\epsilon_2 t(s-s_2)}, \nonumber
\end{eqnarray}
and $s=1/(1-\ep_1/\ep_2)$ and $t=1/(1-\sigma_1/\sigma_2)$ are the dielectric
and conductivity contrasts, respectively.
Note that we predict two characteristic frequencies $f_{c1}$ and $f_{c2}$.

Regarding the dynamic viscosity, the noninteracting spin friction 
expression for an isolated spherical particle must be modified. For many 
particles interacting in a suspension, we must consider the suspension 
hydrodynamics. 
The spin friction is calculated by considering the hydrodynamics of a 
rotating particle in a homogeneous suspension with an effective viscosity 
$\eta_e$. The effective viscosity for a suspension may be obtained 
from an analogous Maxwell-Garnett approximation (MGA) \cite{Choy}:
\be
{\eta_e - \eta_2 \over \eta_e + {3\over 2} \eta_2} 
 = p {\eta_1 - \eta_2 \over \eta_1 + {3\over 2} \eta_2},
\label{eta1}
\de
where $\eta_1$ and $\eta_2$ are respectively the viscocity of particle 
and host, and we may take $\eta_1 \to \infty$ for hard spheres. 
The dilute limit expression for the effective viscocity is:
\be
\eta_e=\eta_2+\left({5\over 2}p\right)\eta_2
 +\left({5\over 2}p^2\right)\eta_2+\cdots.
\label{mgad}
\de

This equation is a suitable method to determine the effective viscosity 
at low concentration as it predicts no percolation threshold.
The effective medium approximation (EMA) is also a useful method to 
evaluate the effective viscosity of a suspension, which admits the form 
\cite{Choy}:

\be
p\frac{\eta_1-\eta_e}{\eta_1+{3\over 2}\eta_e}
 +(1-p)\frac{\eta_2-\eta_e}{\eta_2+{3\over 2}\eta_e}=0.
\label{eta2}
\de
We should remark that a percolation threshold $p_c=0.4$ is predicted by 
Eq.(\ref{eta2}).
From this equation, the dilute limit expression is
\be
\eta_e=\eta_2+\left({5\over 2}p\right)\eta_2
 +\left({5\over 2}p\right)^2\eta_2+\cdots.
\label{emad}
\de
To one's interest, Eq.(\ref{mgad}) and Eq.(\ref{emad}) attain the same 
expansion up to first order in $p$. In addition, we should also remark 
that both Eq.(\ref{eta1}) and Eq.(\ref{eta2}) are valid only at low
Reynolds number.

It is known that the rotational angular velocity of a particle is related 
to the dipole factor as follows
\be
\Omega(f)=-\frac{\epsilon_e E_0^2}{2\eta_e} Im[b]
\de
where $E_0$ is the strength of the applied electric field, and 
$Im[\cdots]$ means taking the imaginary part of $[\cdots]$.

\section{Numerical results}

During the numerical calculation, set $E_0=1.0\times 10^4 V/m$, 
$\eta_2=1.0\times 10^{-3} Kg/m\cdot s$. 
Note $\ep_0$ denotes the dielectric constant of free space.
 
Fig.1(a) is plotted versus frequency $f$ for three different $\epsilon_1$ 
at $\sigma_2=2.8\times 10^{-4} S/m$, $\epsilon_2=80\epsilon_0$, $p=0.05$ 
and $\sigma_1=2.8\times 10^{-2} S/m$. 
Given $\epsilon_1$, there is always one peak for angular velocity as 
frequency increases. The frequency 
located by the peak is just the characteristic frequency. We should 
remark here that two peaks we predict 
above are located so close that only one peak behavior is shown. In our 
previous work \cite{Huang2}, a pair 
of particles is investigated by considering the mutual depolarization 
effect, and we found that one peak is 
dominant while the other has a lower value and is located in a quite 
different position. This shows 
that the particle aggregation effect may enhance the second peak value, 
and be relocated close to the dominant 
one. Regarding the related quantitative explanation, please refer to 
Fig.3 arising from the spectral 
representation approach. As $\ep_1$ increases, both the characteristic 
frequency and the angular-velocity peak value decreases.

Fig.1(b) is plotted versus frequency $f$ for three different $\epsilon_2$ 
at $\sigma_2=2.8\times 10^{-4} S/m$, $p=0.05$, $\epsilon_1=10\epsilon_0$ 
and $\sigma_1=2.8\times 10^{-2} S/m$. 
It is evident that increasing $\ep_2$ yields decreasing characteristic 
frequency, but increasing angular-velocity peak value.

Fig.1(c) is plotted versus frequency $f$ for three different $\sigma_1$ 
at $\sigma_2=2.8\times 10^{-4} S/m$, $\epsilon_2=80\epsilon_0$, $p=0.05$ 
and $\epsilon_1=10\epsilon_0$. 
It is evident that both the characteristic frequency and the 
angular-velocity peak value increase for increasing $\sigma_1$.

Fig.1(d) is plotted versus frequency $f$ for three different $\sigma_2$ 
at $\epsilon_2=80\epsilon_0$, $p=0.05$ and $\epsilon_1=10\epsilon_0$ and 
$\sigma_1=2.8\times 10^{-2} S/m$. 
For an increasing $\sigma_2$, the characteristic frequency increases 
whereas the angular-velocity peak value decreases concomitantly.

In Fig.2, the effect of volume fraction on the effective viscosity is 
investigated. 
Clearly, at a larger volume fraction, a larger effective viscosity is 
predicted both for MGA as well as EMA. 
It is evident that, for a given $p$, EMA predicts a larger effective 
viscosity than MGA.

In Fig.3, we investigate the dependence of the spectral parameters, 
namely the residues ($F_1$ and $F_2$), the poles ($s_1$ and $s_2$), 
as well as the dispersion parameters ($\Delta \epsilon_1$ and 
$\Delta \epsilon_2$) on the volume fraction $p$ (for MGA only). 

In Fig.3(a) and 3(b), as the volume fraction $p$ increases, 
both $F_1$ and $F_2$ increase. Increasing volume fraction $p$ 
leads to decreasing $s_1$, but increasing $s_2$. 
Note that the residues satisfy the relation $F_1+F_2=-(1-p)/3$.  
In the two figures, the distinction between $F_1$ and $F_2$, or
$s_1$ and $s_2$ is quite small at low volume fraction (e.g., $p=0.05$, 
$0.1$, and $0.15$). 

In Fig.3(c), as the volume fraction increases, $\Delta \epsilon_1$ first 
increases, and then decreases, 
while $\Delta\epsilon_2$ decreases monotonically. Note $\Delta 
\epsilon_2$ becomes very close 
to $\Delta \epsilon_1$ at low concentration region. 

In Fig.3(d), as the volume fraction increases, $f_{c1}$ decreases, 
while $f_{c2}$ increases concomitantly. 
Moreover, it is evident that both characteristic frequencies collapse 
at low concentration region.

From Fig.3, it is readily concluded that at low volume fraction, 
the angular-velocity peaks related to two characteristic frequencies 
are so close that they can almost be seen as overlapped. 
This is just the reason why only one rotation peak occurs in our figures. 
We believe that such distinction between the two peaks
will become recognizable for a high-concentration case. 

Fig.4 is plotted for three different volume fraction $p$ at 
$\sigma_2=2.8\times 10^{-4} S/m$, $\epsilon_2=80\epsilon_0$, and 
$\epsilon_1=10\epsilon_0$ and 
$\sigma_1=2.8\times 10^{-2} S/m$. In this figure, we investigate the 
effect of particle volume fraction on the angular velocity. Increasing 
volume fraction is able to reduce the characteristic frequency as well 
 as the angular-velocity peak value. 

\section{DISCUSSION AND CONCLUSION}

Here a few comments on our results are in order. We considered 
homogeneous, 
spherical particles exposed to a rotating electric field. 
As the strength of the rotating electric field increases, the particles 
in a suspension may be aggregated into sheet-like structures. 
Within these structures, the concentration of particles is high. 
In this case, the local-field effect arising from the many-particle 
system must be taken into account. In this regard, we believe the 
effective medium theories including both the Maxwell-Garnett 
approximation (also known as the Clausis-Mossotti approximation) and 
the effective medium approximation (also known as the Bruggeman
approximation) are good candidate theories.
The Maxwell-Garnett approximation is well-known to be non-symmetrical and 
May thus be suitable for low concentration. For a higher concentration 
of particles, we had better use the effective medium approximation (EMA). 
Both methods are valid at low concentrations whereas EMA predicts a 
percolation threshold \cite{Choy}.

We believe our theory will be valid for the case of low Reynolds numbers.
In our present study, both the radii and the angular velocities of the 
particles are small. 
Thus, we may safely use the effective medium theories for the viscosity, 
which are valid when the particles are at rest.

We realize the appearance of a circular medium flow because all the 
particles rotate in the same direction. The macroscopic spin rate
may drive the suspending liquid, leading to a decrease of the apparent
viscosity of the suspension \cite{Brenner}.
Finally, we can apply our formalism to single or multi shell objects,
like biological cells \cite{Huang}. 

In summary, we have investigated the electrorotation spectrum of many 
interacting particles including their electric and hydrodynamic 
interactions. The results showed that the electrorotation spectrum can 
be modified due to the local field effects arising from the many-particle 
system. The hydrodynamic interactions between the suspended particles can 
also modify the spin friction, which is a key to determine the angular 
velocity of electrorotation. By invoking the spectral representation 
approach, we derived the analytic expressions for the characteristic 
frequency at which the maximum angular velocity of electrorotation 
occurs. 
We reported a detailed investigation of the dependence of the 
characteristic frequency and the dispersion strength on various 
material parameters. 

\section*{Acknowledgments}

This work was supported by the Research Grants Council of the Hong Kong 
SAR Government under grant CUHK 4245/01P. 
G.Q.G. acknowledges the financial support from a Key Project of the 
National Natural Science Foundation of China under grant 19834070.
The final version of this paper was completed during the Forum on 
Condensed Matter and Interdisciplinary Physics held at Nanjing University,
China on the occasion of the centennial anniversary of the University. 
K.W.Y. thanks the Organizing Committee for their invitation and the 
hospitality received from the Laboratory of Solid State Microstructures 
and the Department of Physics during his stay at the Nanjing University.

\newpage

\begin{figure}[h]
\caption{Dependence of rotational 
angular velocity on $\ep_1$, $\ep_2$, $\sigma_1$ and $\sigma_2$ versus 
frequency (for the MGA only).
(a) $\sigma_2=2.8\times 10^{-4} S/m$, $\epsilon_2=80\epsilon_0$, $p=0.05$ 
and $\sigma_1=2.8\times 10^{-2} S/m$. 
(b) $\sigma_2=2.8\times 10^{-4} S/m$, $p=0.05$ and 
$\epsilon_1=10\epsilon_0$ and $\sigma_1=2.8\times 10^{-2} S/m$.
(c) $\sigma_2=2.8\times 10^{-4} S/m$, $\epsilon_2=80\epsilon_0$, $p=0.05$ 
and $\epsilon_1=10\epsilon_0$.
(d) $\epsilon_2=80\epsilon_0$, $p=0.05$ and $\epsilon_1=10\epsilon_0$ and 
$\sigma_1=2.8\times 10^{-2} S/m$. 
}
\end{figure}

\begin{figure}[h]
\caption{The MGA and EMA effective viscosity of suspension plotted versus 
volume fraction $p$.}
\end{figure}

\begin{figure}[h]
\caption{The spectral parameters $F_1$, $F_2$, $s_1$ and $s_2$, 
as well as the dispersion parameters, $\Delta \epsilon_1$ and 
$\Delta \epsilon_2$, plotted versus volume fraction $p$ (for the MGA only).}
\end{figure}

\begin{figure}[h]
\caption{The rotational angular velocity plotted versus frequency for 
different volume fraction $p$. Parameters used in the plot are
$\epsilon_1=10\epsilon_0$, 
$\sigma_1=2.8\times 10^{-2} S/m$,
$\sigma_2=2.8\times 10^{-4} S/m$, $\epsilon_2=80\epsilon_0$}
\end{figure}

\newpage
\centerline{\epsfig{file=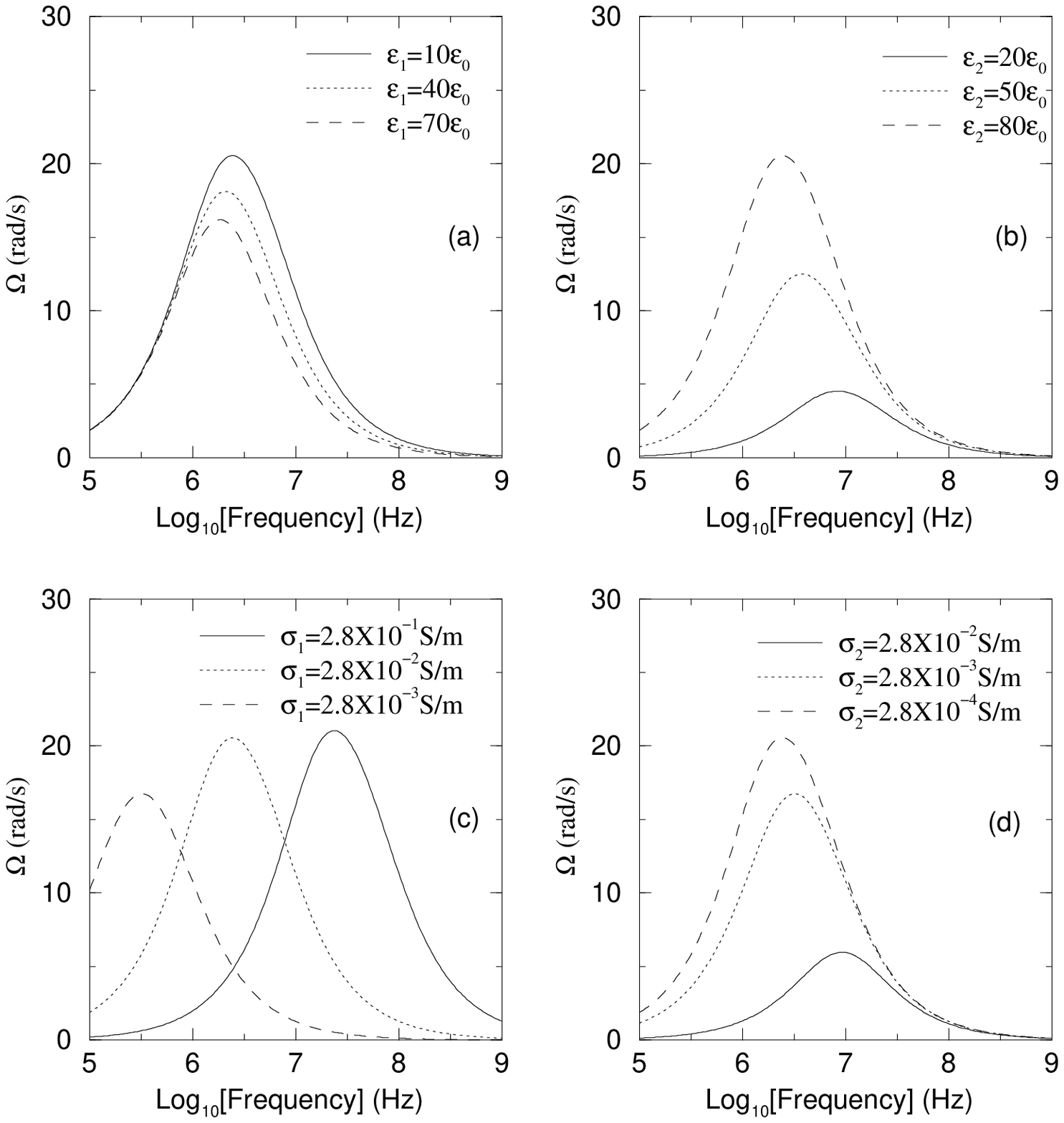,width=\linewidth}}
\centerline{Fig.1}

\centerline{\epsfig{file=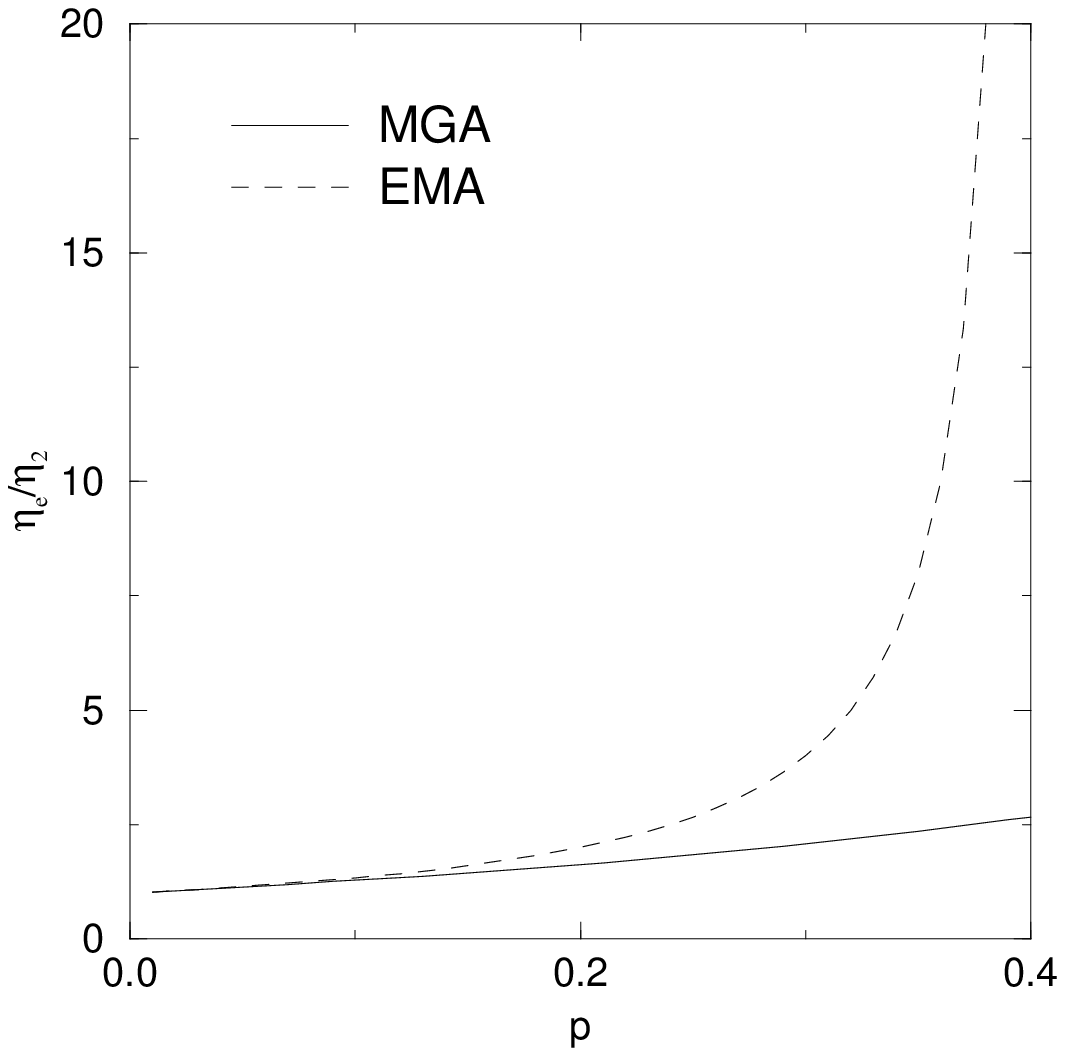,width=\linewidth}}
\centerline{Fig.2}

\centerline{\epsfig{file=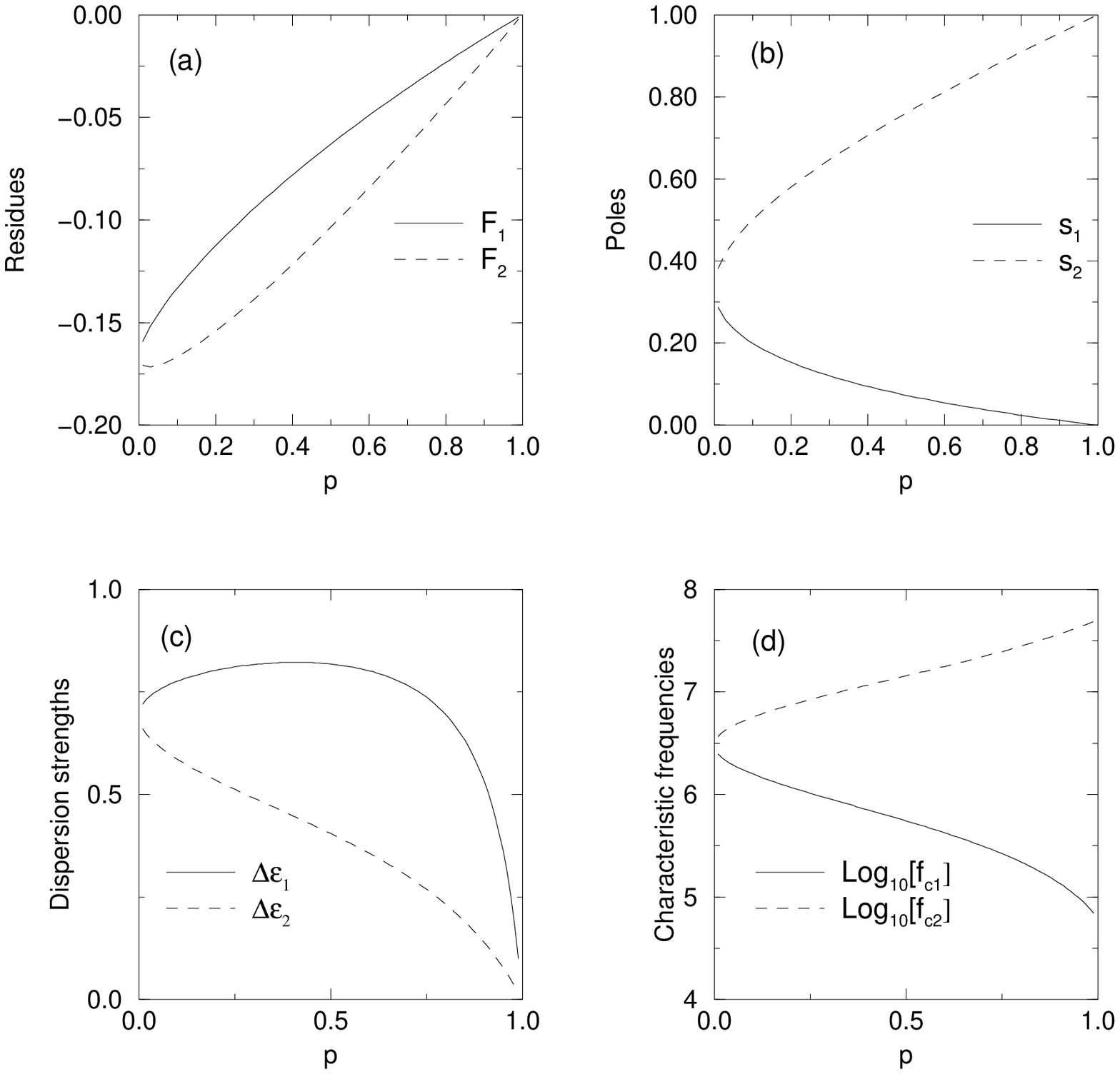,width=\linewidth}}
\centerline{Fig.3}

\centerline{\epsfig{file=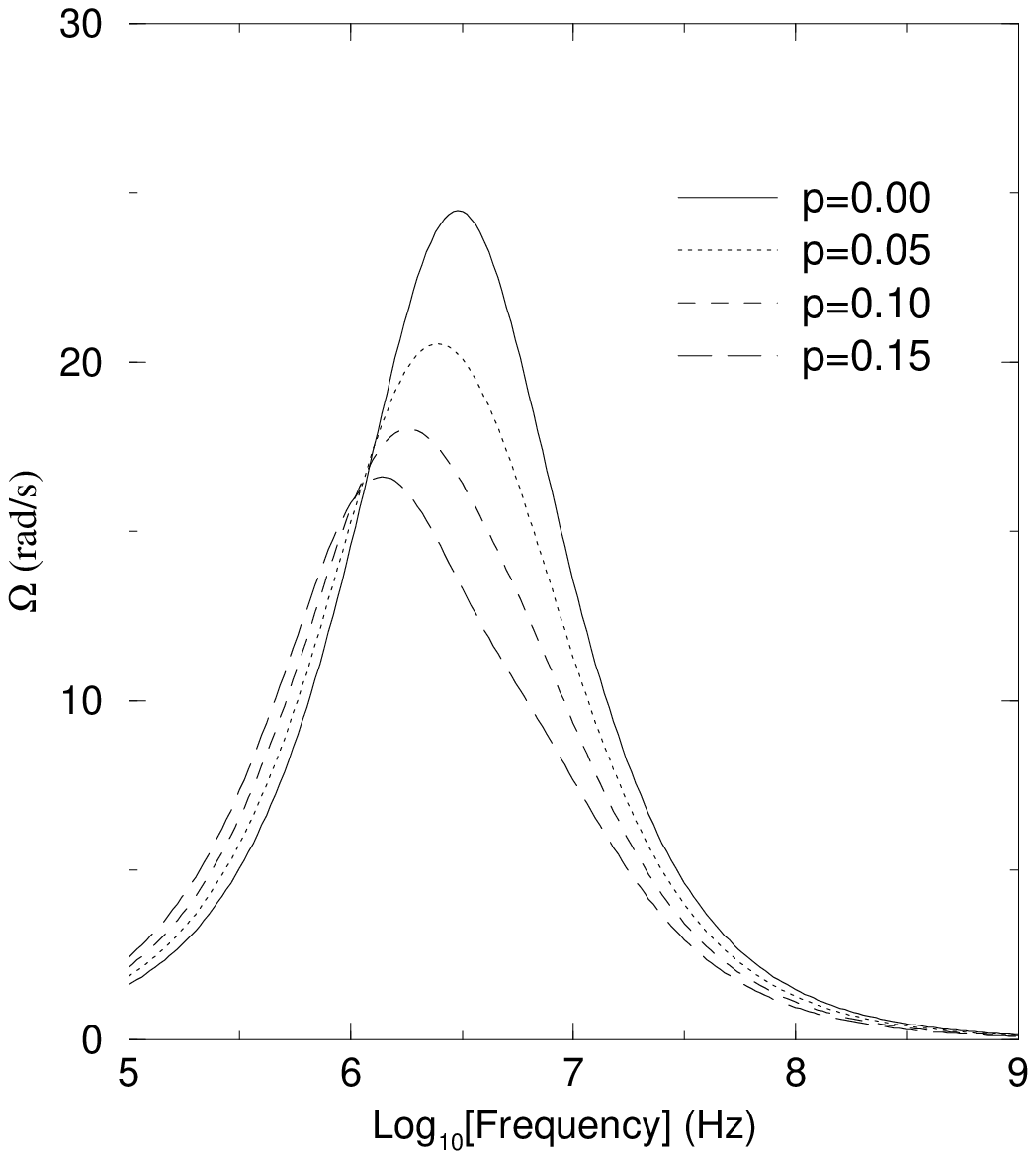,width=\linewidth}}
\centerline{Fig.4}

\end{document}